\documentstyle[epsfig,12pt]{article}

\begin{document}

\begin{titlepage}

\vspace*{2.truecm}

\centerline{\large \bf Phase Ordering Dynamics of $\phi^4$ Theory}
\vskip 0.6truecm
\centerline{\large \bf with Hamiltonian Equations of Motion}
\vskip 0.6truecm

\vskip 2.0truecm
\centerline{\bf B. Zheng$^{1,2}$, V. Linke$^3$ and S. Trimper$^1$}
\vskip 0.2truecm

\vskip 0.2truecm
\centerline{$^1$ FB Physik, Universit\"at Halle, 06099 Halle, Germany}
\vskip 0.2truecm
\centerline{$^2$ ICTP, 34014 Trieste, Italy}
\vskip 0.2truecm
\centerline{$^3$ FB Physik, Freie Universit\"at Berlin, 14195 Berlin, Germany}
  
\vskip 2.5truecm

\abstract{Phase ordering dynamics of the $(2+1)$- and
$(3+1)$-dimensional $\phi^4$ theory
with Hamiltonian equations of motion is investigated numerically.
Dynamic scaling is confirmed. The dynamic exponent $z$ is 
different from that of the Ising model with dynamics of model A,
while the exponent $\lambda$ is the same.
}

\vspace{0.5cm}



\end{titlepage}

\section{Introduction}

It is believed that macroscopic properties of many particle
systems could be, in principle, 
 described by microscopic deterministic equations of motion
(e.g., Newton, Hamiltonian and Heisenberg equations),
if all interactions, boundary conditions and initial
states could be taken into account. However, practically 
it is very difficult to solve these equations,
except for some simple cases. Therefore,
statistical mechanics is developed to describe effectively 
statistical systems.
In usual circumstances, ensemble theories give good
description of equilibrium states.
But it is complicated for non-equilibrium states.
A general theory does not exist. In many cases,
some stochastic dynamics, 
e.g., that described by Langevin-type equations of motion 
or Monte Carlo dynamics,
is considered to be approximate theories.
Anyway, whether microscopic equations of motion 
could really produce the
results of statistical mechanics, or vice verse,
remains open, e.g., see Refs~ \cite {fer65,for92,esc94,ant95,cai98}.

With recent development of computers,
gradually it becomes possible 
to solve microscopic deterministic equations {\it numerically}. 
This attracts much attention of scientists
in different fields. Study of microscopic fundamental dynamics
is on the one hand to test statistical mechanics,
 and on the other hand to explore new physics. 
For example, assuming a system is isolated,
there is only internal interaction, and
periodic boundary conditions can be adopted.
Computations are greatly simplified.
To achieve ergodicity, the system should start from
random initial states. 
Recently, such effort has been made 
for the $O(N)$ vector model
and $XY$ model \cite {cai98,cai98a,leo98}. 
The results support that
 deterministic Hamiltonian equations correctly describe
second order phase transitions.
The estimated static critical exponents are consistent with
those calculated from canonical ensembles.
More interestingly, the macroscopic short-time (non-equilibrium)
dynamic behavior of the $(2+1)$-dimensional
$\phi^4$ theory at {\it criticality} has also been investigated 
and dynamic scaling is found
\cite {zhe99,zhe99a}.
The results indicate that Hamiltonian dynamics 
in two dimensions with random initial states
is in a same universality class of 
Monte Carlo dynamics of model~A. 

In a similar spirit, in Ref. \cite {zhe00b} the author
has investigated phase ordering dynamics
of the $(2+1)$-dimensional 
 $\phi^4$ theory with Hamiltonian equations of motion.
 Assuming random initial states,
 there is a minimum energy density
 which is above the  real minimum energy density of the system.
   Starting from this minimum energy density
    (noting that energy is conserved),
    which is well below the critical energy density, 
 phase ordering occurs. Dynamic scaling behavior is found.
 The dynamic exponent $z$ is different from that of
 model A dynamics, but the exponent $\lambda$ governing
 the power law decay of the autocorrelation looks the same.
 Somewhat interesting is that the scaling function 
 of the equal-time spatial correlation function
 is the same as that of the Ising model with model A dynamics.
 All above results are independent of the parameters
 in the system.

The purpose of this paper is twofold:
Firstly, we generalize the computations
to $(3+1)$ dimensions. This is important since our realistic world is
in $(3+1)$ dimensions. Furthermore, in phase ordering of model A dynamics,
the dynamic exponent $z$ is dimension independent
but the exponent $\lambda$ is dimension dependent.
It is interesting whether this property is kept in
Hamiltonian dynamics. Attention will be also put on
whether the scaling function of the equal-time spatial
 correlation function in three dimensions is the same
 as that of the Ising model with model A dynamics.
Secondly, to achieve more confidence on our conclusions
we will reexamine the results for $(2+1)$ dimensions
in Ref. \cite {zhe00b} using somewhat different, more careful approaches.
Since the computations in $(3+1)$ dimensions
are very much time consuming, more accurate data are obtained
in $(2+1)$ dimensions.

In the next section, we introduce the model and analyze
the dynamic scaling behavior. In Sec. 3, numerical results are
presented. Finally come the concluding remarks.

\section{Phase ordering dynamics}

In the following, we outline phase ordering dynamics with Hamiltonian 
equations of motion. For a recent review of general ordering dynamics,
readers are referred to Ref. \cite {bra94}.

\subsection{The model}

For an isolated system, the Hamiltonian of the 
$(d+1)$-dimensional $\phi^4$ theory
on a square or cubic lattice is 
\begin{equation}
H=\sum_i \left [ \frac{1}{2} \pi_i^2 
  + \frac{1}{2} \sum_\mu (\phi_{i+\mu}-\phi_i)^2 
  - \frac{1}{2} m^2 \phi_i^2 
  + \frac{1}{4!} g \phi_i^4 \right ]
\label{e10}
\end{equation}
with $\pi_i=\dot \phi_i$. It leads to the equations of motion
\begin{equation}
 \ddot \phi_i= \sum_\mu (\phi_{i+\mu}+\phi_{i-\mu}- 2\phi_i)
  +  m^2 \phi_i
  - \frac{1}{3!} g \phi_i^3\ .
\label{e20}
\end{equation}
Here $\mu$ represents spatial directions.
Energy is conserved in these equations. 
Solutions are assumed to generate a microc-anonical ensemble.
The temperature could be defined as the averaged
kinetic energy. For the {\it non-equilibrium} dynamic system, however,
total energy is a more convenient controlling parameter,
 since it is conserved and 
can be taken as input from initial states. 
For given parameters $m^2$ and $g$,
there exists a critical energy density
$\epsilon_c$, separating the ordered phase 
(below $\epsilon_c$) and disordered phase (above $\epsilon_c$).
The phase transition is of second order.

We should emphasize that a Langevin equation at zero temperature
is also `deterministic' in the sense that
there are no noises, but it is essentially different from the
Hamiltonian equation (\ref {e20}). 
The former describes relaxation towards equilibrium at zero temperature
for a non-isolated system,
but the latter contains full physics at all temperatures
for an isolated system.

The order parameter of the $\phi^4$ theory is the magnetization.
The time-dependent magnetization $M\equiv M^{(1)}(t)$  and its second moment
 $M^{(2)}$ are defined as
\begin{equation}
M^{(k)}(t)=\frac {1}{L^{dk}} 
\langle \left [ \sum_i \phi_i(t) \right ]^{(k)} \rangle, \quad k=1, 2.
\label{e30}
\end{equation}
$L$ is the lattice size and $d$ is the spatial dimension. 
Here it is important that the average is {\it 
over initial configurations}. This is a real sample
average and different from the time average in equilibrium.

Following ordering dynamics with stochastic equations
\cite {bra94},
we consider a dynamic process that the system, initially in
a {\it disordered} state but with an energy density
well below $\epsilon_c$,
is suddenly released to evolve according to Eq.~(\ref {e20}).
For simplicity,
we set initial kinetic energy to 
zero, i.e., $\dot \phi_i(0)=0$.
To generate a random initial configuration $\{\phi_i(0)\}$,
we first fix the magnitude  
$|\phi_i(0)| \equiv c$, then randomly give the sign to $\phi_i(0)$
with the restriction of a fixed magnetization in units of $c$,
and finally the constant $c$ is determined by
the given energy.

In case of stochastic dynamics,
 scaling behavior of phase ordering is dominated
by the fixed point $(T_I,T_F)=(\infty,0)$
with $T_I$ being the initial temperature
and $T_F$ being the temperature after quenching \cite {bra94}.
In Hamiltonian dynamics,
the energy density can not be taken to the real minimum
$e_{min} = - 3 m^4/2 g$  
since the system does not move.
Actually, for the initial states described above,
the energy is given by
\begin{equation}
V=\sum_i \left [ (d - \frac{1}{2} m^2) \phi_i^2 
  + \frac{1}{4!} g \phi_i^4 \right ] \ .
\label{e40}
\end{equation}
For the case of $d < m^2/2$, it is demonstrated
in Ref. \cite {zhe00b} that for a energy density
well below the critical point $\epsilon_c$, due to the competition of two
ordered states, phase ordering occurs 
when the initial magnetization is set to zero.
The scaling behavior
is dominated by the minimum energy density $v_{min}=V_{min}/L^d$,
which is a kind of fixed points.
Above $v_{min}$, there are extra corrections to scaling.
From now, we redefine the energy density $e_{min}$ 
as zero. Then the fixed point is $\epsilon_0=v_{min}-e_{min}$.
In this paper, we consider only the energy density
at exactly the fixed point $\epsilon_0$. 

\subsection{Dynamic scaling behavior}

Let us first consider
the case of the initial magnetization $m_0=0$.
An important observable is
the equal-time correlation function 
\begin{equation}
C(r,t) =\frac {1}{L^{d}} 
\langle \sum_i \phi_i (t) \phi_{i+r} (t) \rangle \ .
\label{e50}
\end{equation}
The scaling hypothesis is that 
at the {\it late} stage of the time evolution,
$C(r,t)$ obeys a scaling form
\begin{equation}
C(r,t) = f(r/t^{1/z}) \ ,
\label{e60}
\end{equation}
where $z$ is the so-called dynamic exponent.
To the understanding of the authors, 
here 'late' is meant in microscopic sense. In other words,
when the domain size ($\sim t^{1/z}$) is big enough 
in units of the lattice spacing, scaling behavior
emerges. At finite $t$, of course, 
there may be corrections to scaling.
Corrections to scaling are generally not universal.
They may induce difficulties for observing
scaling behavior and uncertainties in the determination
of the critical exponents. 

Simple understanding of the scaling behavior of
$C(r,t)$ can be achieved from the second moment
of the magnetization.
Integrating
over $r$ in Eq.~(\ref {e60}), we obtain a power law behavior 
\begin{equation}
M^{(2)}(t) \sim t^{d/z} \ .
\label{e70}
\end{equation}

Another interesting observable is the auto-correlation function
\begin{equation}
A(t)=\frac {1}{L^d}\langle \sum_i  \phi_i(0) \phi_i(t) \rangle  .
\label{e80}
\end{equation}
The scaling hypothesis
leads to a power law behavior 
\begin{equation}
A(t) \sim t^{-\lambda/z},
\label{e90}
\end{equation}
which implies that ordering dynamics is in some sense {\it `critical'}.
Here $\lambda$ is another independent exponent. 

For the discussions above, the initial magnetization $m_0$
is set to zero. If $m_0$ is non-zero,
the system reaches a unique ordered state within a finite time.
If $m_0$ is infinitesimally small, however,
the time for reaching the ordered state is also infinite
and scaling behavior can still be expected, at least 
at relatively early times (in macroscopic sense).
In this case, an interesting observable is the magnetization itself.
It increases by a power law
\begin{equation}
M(t) \sim t^{\theta}, \quad \theta=(d-\lambda)/z.
\label{e100}
\end{equation}
The exponent $\theta$ can be written as $x_0/z$, with $x_0$
being the scaling dimension of $m_0$.
This power law behavior has deeply been investigated
in critical dynamics \cite {jan89,zhe98}.
The interesting point here is that $\theta$
is related to the exponent $\lambda$ which governs the power law decay
of the auto-correlation. By combining measurements of 
$\theta$ and $\lambda$, one can also estimate the dynamic exponent $z$. 

\section{Numerical results}

To solve the equations of motion (\ref {e20}) numerically,
 we discretize
$\ddot \phi_i$ by 
$(\phi_i(t+\Delta t)+\phi_i(t-\Delta t)-2\phi_i(t))/(\Delta t)^2$.
Starting from an initial configuration,
we update the equations of motion up to
a certain maximum time $t_{max}$.
 Then we repeat the procedure with other
initial configurations.
In Ref. \cite {zhe00b}, reasonable results 
in two dimensions are obtained mainly with
$\Delta t=0.05$ up to $t_{max}=640$ and a lattice size
$L=521$. $200$ samples of initial configurations
 are used for averaging.
For three dimensions, we also perform the computations
with $\Delta t=0.05$ up to $t_{max}=640$ but with
a lattice size $L=125$. $50$ samples are taken for averaging.
We have also carried out some computations
with other $\Delta t$'s and lattice sizes to confirm the results.
At the time $t_{max}$, the equal-time correlation function $C(r,t)$
decays to nearly zero at $r \sim 45$ and this indicates
also that the finite size effect with $L=128$ is already small.
Furthermore, in order to gain more confidence in
our conclusions, especially whether our $t_{max}$ has
really reached the scaling regime, 
we perform more accurate computations in two dimensions
(compared with those in Ref. \cite {zhe00b})
with a lattice size $L=256$ and $\Delta t=0.02$, $0.01$
up to $t_{max}=1280$. The number of samples for averaging is $600$.
 Somewhat different and 
more careful approaches will be adopted in this paper.

\begin{figure}[t]
\epsfysize=6.5cm
\epsfclipoff
\fboxsep=0pt
\setlength{\unitlength}{0.6cm}
\begin{picture}(13.6,12)(0,0)
\put(-2.,0){{\epsffile{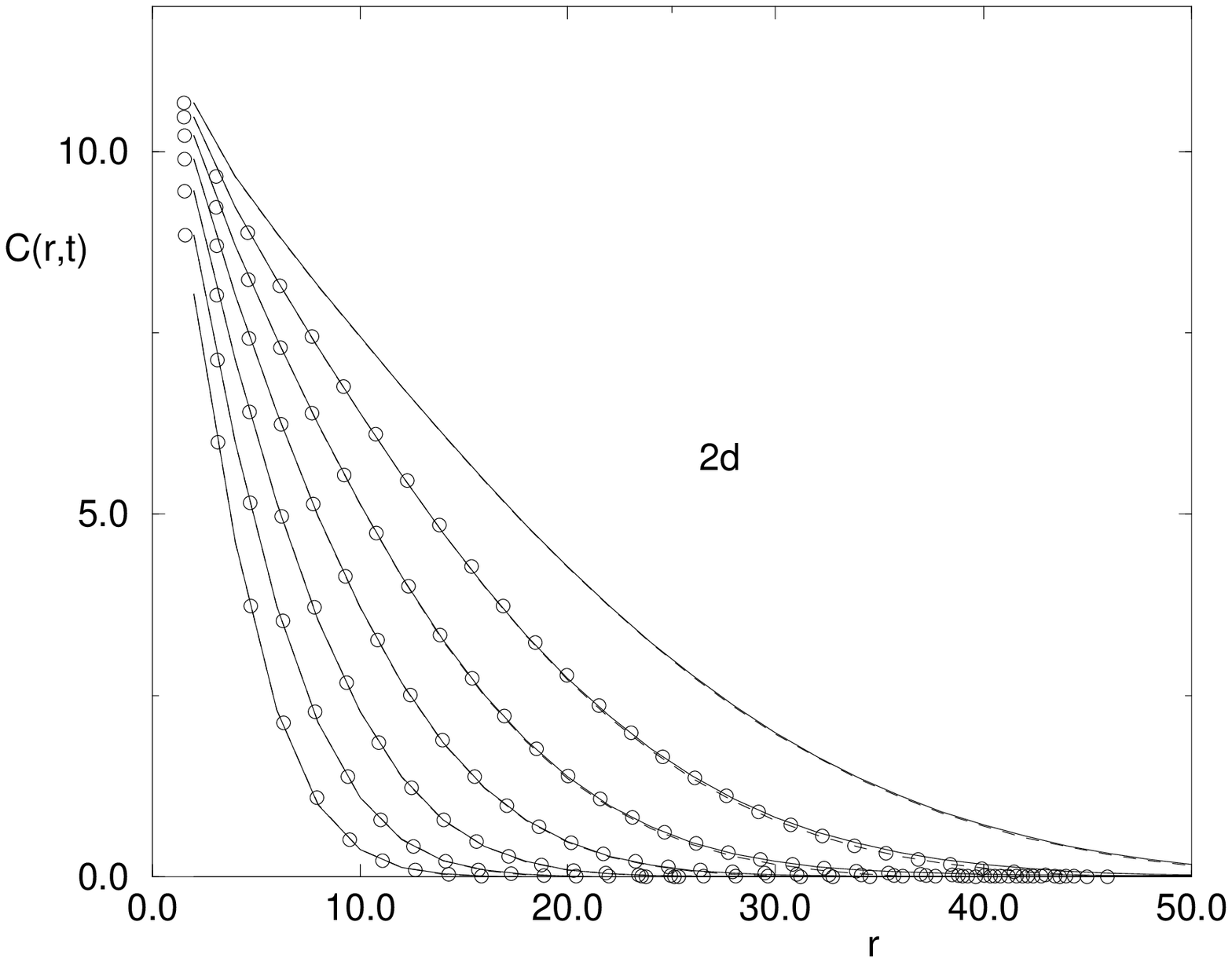}}}
\put(0,-0.0){\makebox(0,0){$(a)$}}
\epsfysize=6.5cm
\put(11.5,0){{\epsffile{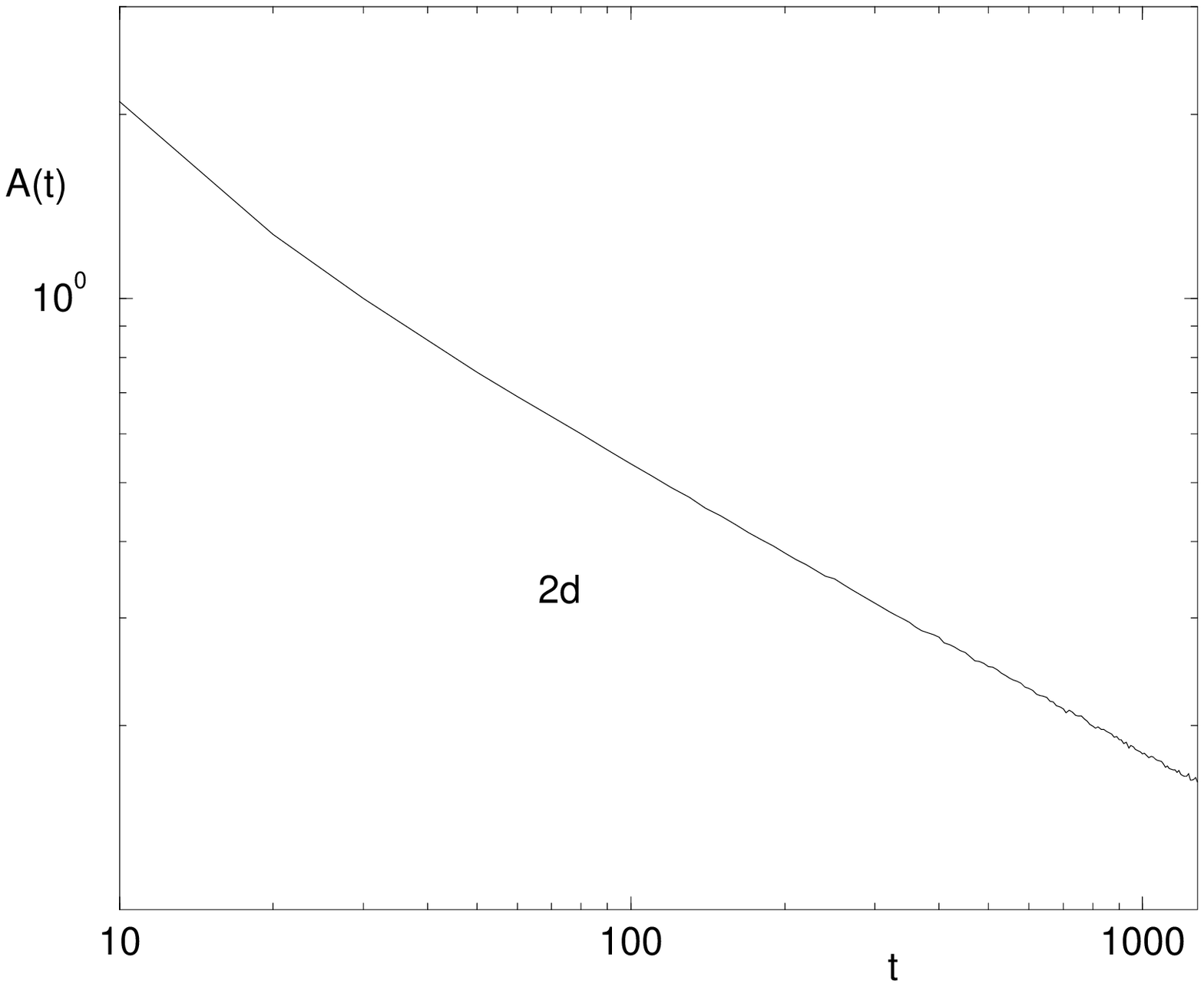}}}
\put(13.5,-0.0){\makebox(0,0){$(b)$}}
\end{picture}
\caption{(a) $C(r,t)$ in two dimensions obtained with $L=256$
and $\Delta t =0.01$ is plotted with solid lines for
$t=20$, $40$, $80$, $160$, $320$, $640$ and $1280$ (from left).
Circles fitted to the curve at the time $t$ are the data
at the time $2 t$ but $r$ being rescaled by a factor $2^{-1/z}$.
$C(r,t)$ obtained with $\Delta t =0.02$ is also plotted with
dashed lines in the figure but they overlap almost completely with 
the solid lines. 
(b) $A(t)$ obtained with $\Delta t =0.01$ (the solid line) in log-log scale.
The curve for $\Delta t =0.02$ overlaps completely
with that for $\Delta t =0.01$.
}
\label{f1}
\end{figure}

In Fig. \ref {f1} (a), the equal-time correlation function
$C(r,t)$ in two dimensions is displayed. Solid lines 
are obtained with $\Delta t=0.01$ and from left
the time $t$ is $20$, $40$, $80$, $160$, $320$, $640$ and
$1280$. Data for $\Delta t=0.02$ are also plotted with 
dashed lines in the figures, but they almost completely
overlap with the solid lines. For the curve of $t=1280$,
$C(r,t)$ decays to nearly zero at $r \sim 50$. 
Therefore, we conclude that
the finite size effect with the lattice size $L=256$
should be already negligible small.
To confirm this,
we have also compared the data with those in Ref. \cite {zhe00b}.
 On the other hand, 
our data show that the finite $\Delta t$
effect for $\Delta t=0.05$ is also negligible.
 According to the scaling form (\ref {e60}),
from data collapse of $C(r,t)$ at different $t$'s
one can estimate the dynamic exponent $z$.
As is observed in Ref. \cite {zhe00b}, 
the effective dynamic exponent $z(t)$ shows a small dependence
on the time $t$. To explore this behavior and extract 
confidently the value of $z$, we perform scaling collapse of
$C(r,t)$ with the time $t$ and $2t$.
In Fig. \ref {f1} (a), circles fitted to a solid line
of the time $t$ are the data of the time $2 t$
with $r$ being rescaled by a factor $2^{-1/z}$,
i.e., $C(r,t)=C(r 2^{1/z},2t)$.
The dynamic exponent $z(t)$
is determined by the best fitting of the circles
to the corresponding solid line.
From the figure, we see clearly that the data collapse
very nicely.

\begin{figure}[t]
\epsfysize=6.5cm
\epsfclipoff
\fboxsep=0pt
\setlength{\unitlength}{0.6cm}
\begin{picture}(13.6,12)(0,0)
\put(-2.,0){{\epsffile{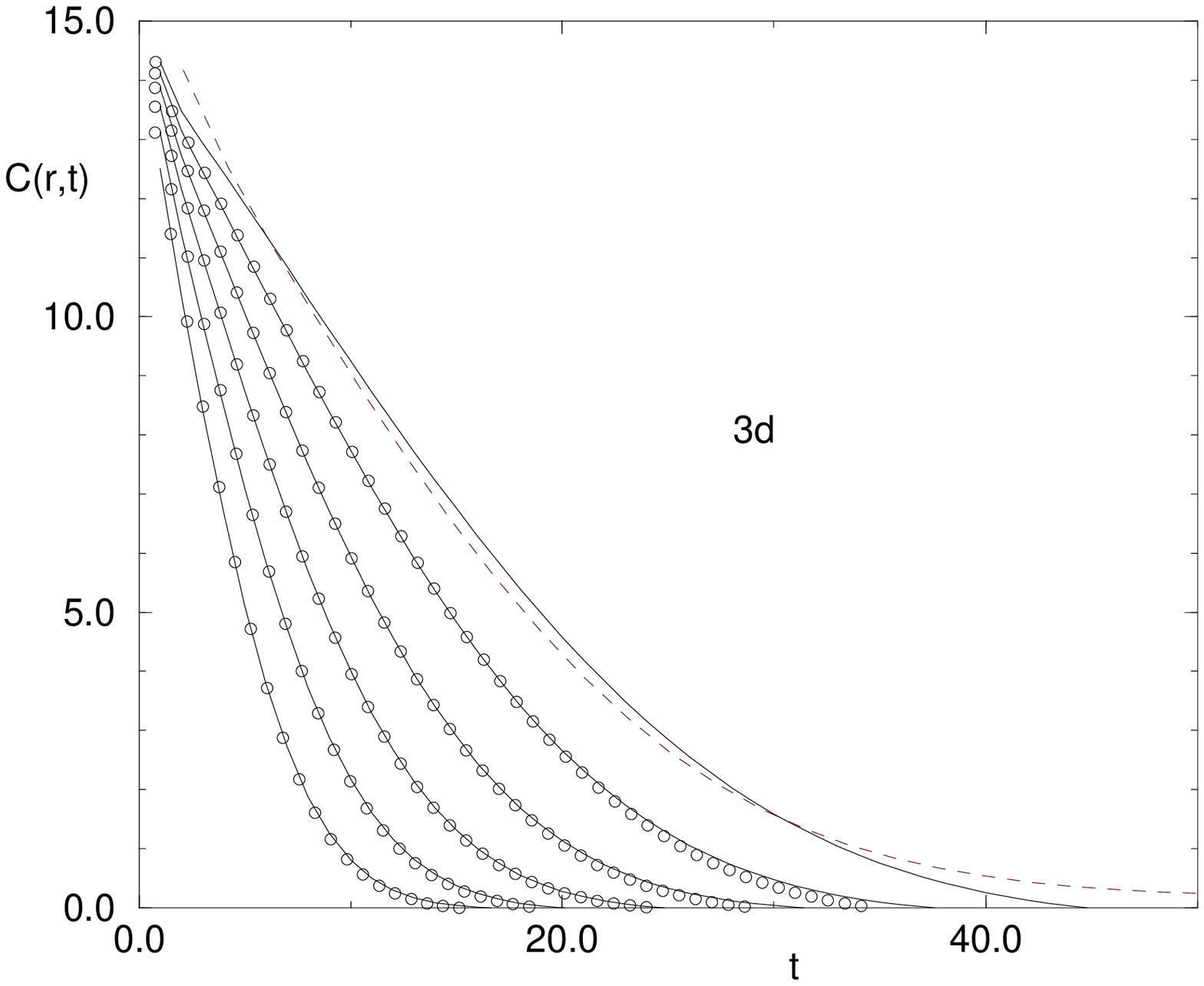}}}
\put(0,-0.){\makebox(0,0){$(a)$}}
\epsfysize=6.5cm
\put(11.5,0){{\epsffile{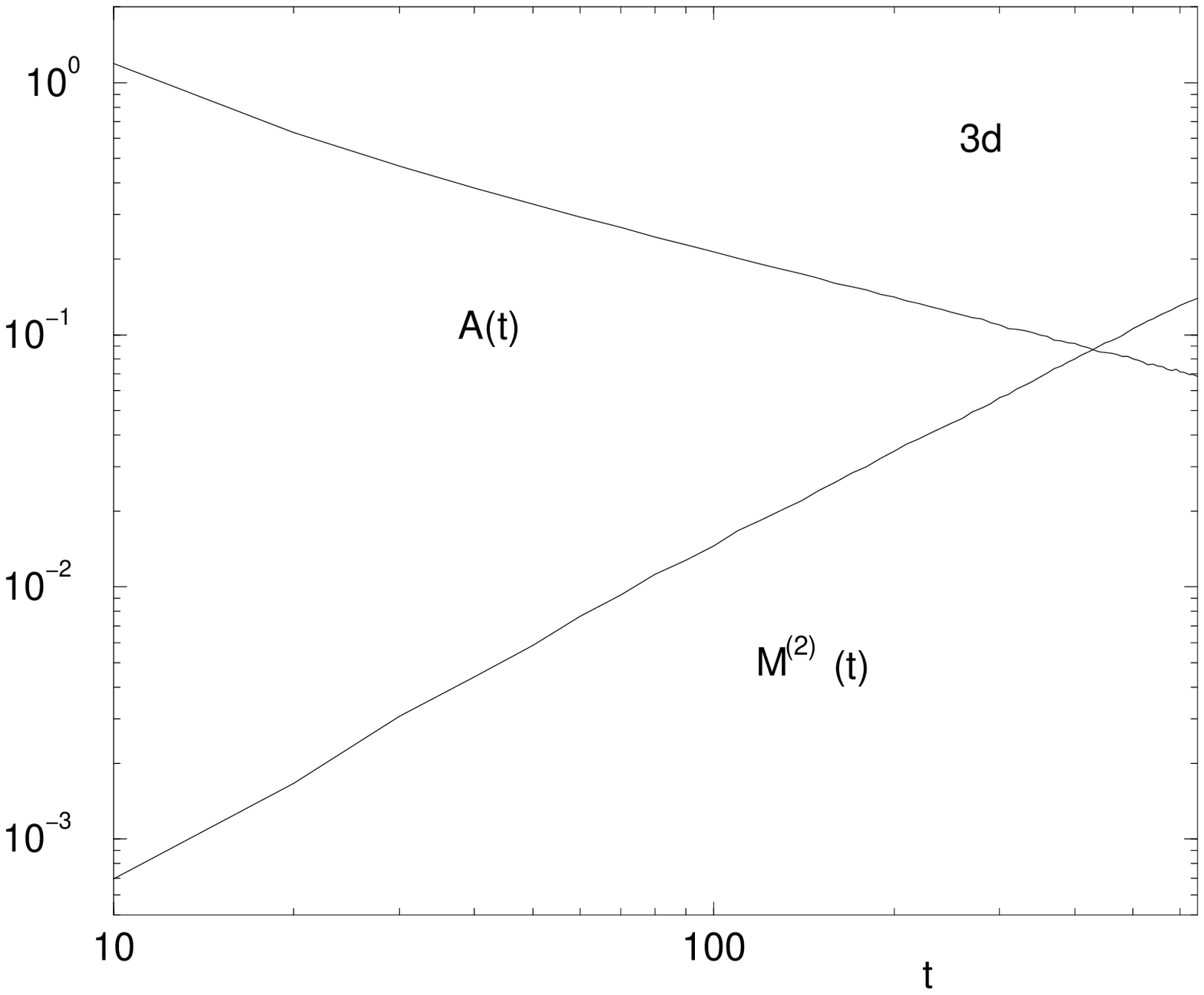}}}
\put(13.5,-0.){\makebox(0,0){$(b)$}}
\end{picture}
\caption{a) $C(r,t)$ in three dimensions obtained with $L=128$
and $\Delta t =0.05$ is plotted with solid lines for
$t=20$, $40$, $80$, $160$, $320$ and $640$ (from left).
Circles fitted to the curve at the time $t$ are the data
at the time $2 t$ but $r$ being rescaled by a factor $2^{-1/z}$.
The dashed line represents the scaling function 
in two dimensions.
(b) $A(t)$ and $M^{(2)}(t)$ in log-log scale.
}
\label{f2}
\end{figure}

\begin{figure}[t]
\epsfysize=6.5cm
\epsfclipoff
\fboxsep=0pt
\setlength{\unitlength}{0.6cm}
\begin{picture}(13.6,12)(0,0)
\put(-2.,0){{\epsffile{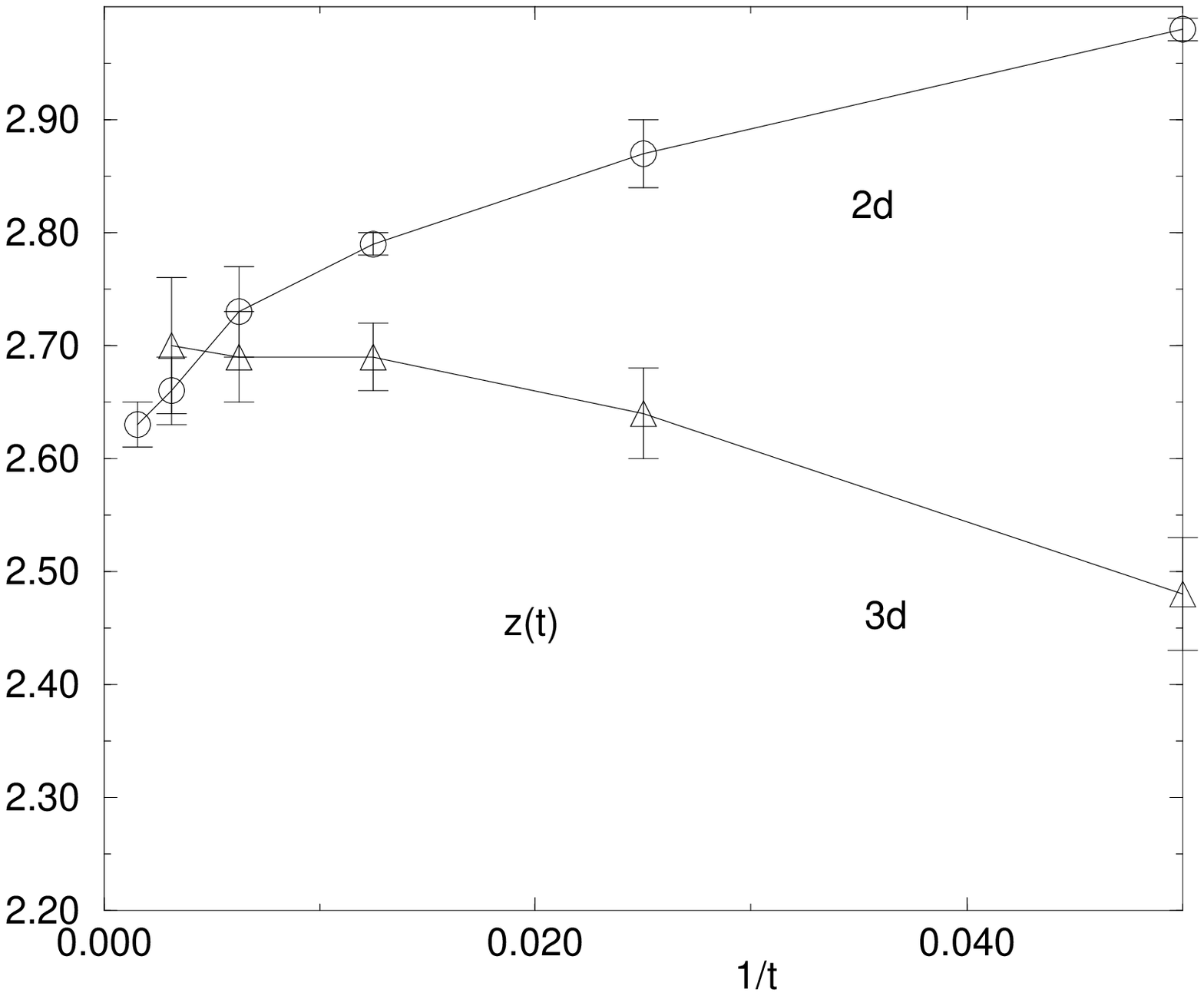}}}
\put(0,-0.){\makebox(0,0){$(a)$}}
\epsfysize=6.5cm
\put(13.5,-0.){\makebox(0,0){$(b)$}}
\put(11.5,0){{\epsffile{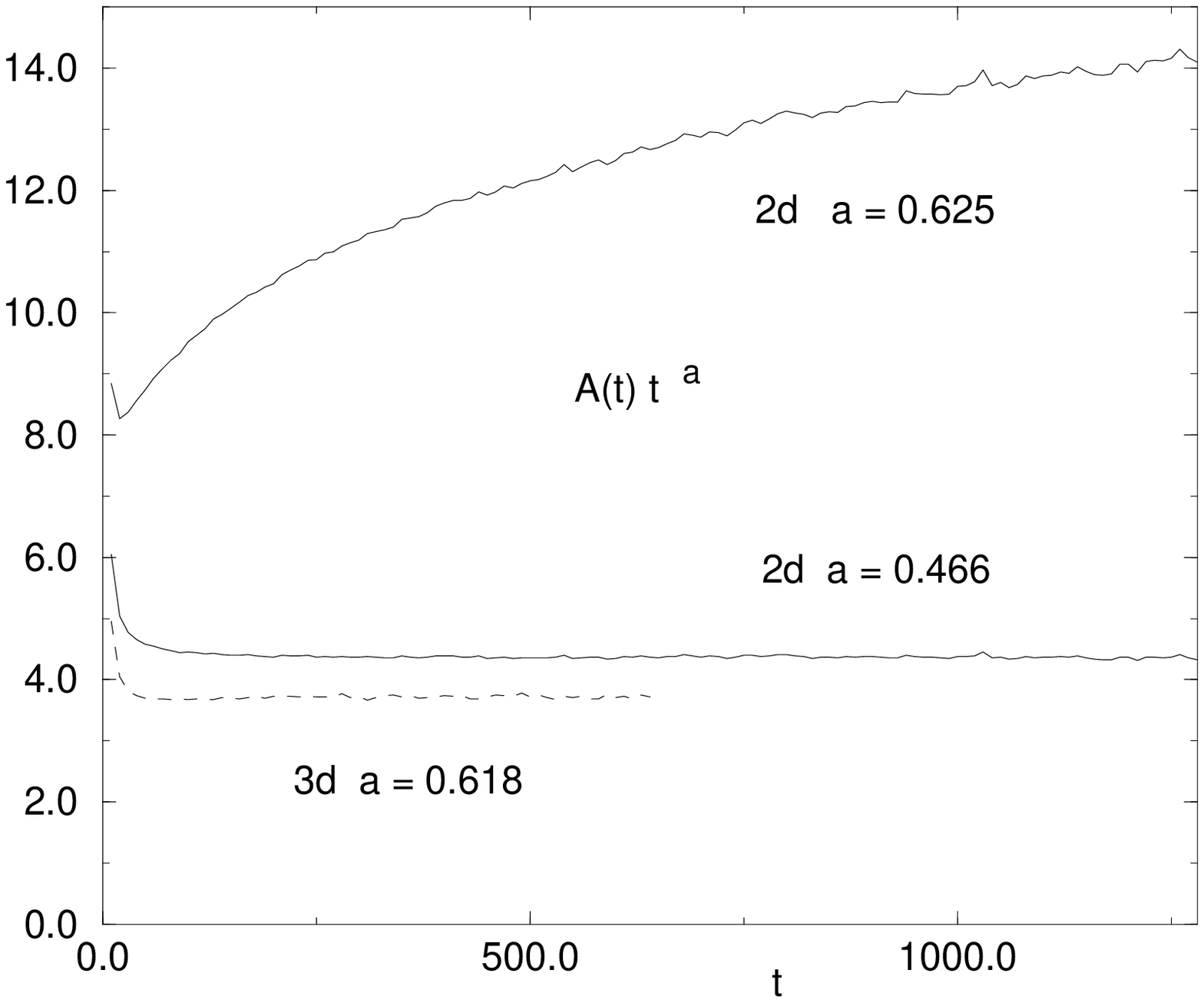}}}
\end{picture}
\caption{(a) The effective dynamic exponent $z(t)$ measured from
scaling collapse of $C(r,t)$ with the times $t$ and $2 t$.
(b) Taking $a=\lambda/z$, $A(t) t^a$ tends to
a constant.
}
\label{f3}
\end{figure}

In Fig. \ref {f2} (a), a similar figure as Fig. \ref {f1} (a)
is shown for $C(r,t)$ in three dimensions. 
Scaling collapse is also observed, even though
for larger $r$ it is not as good as in two dimensions.
This can be neither a finite size effect nor
a finite $\Delta t$ effect, since it exists also for small
$t$'s. 
To see the trend of $z(t)$ as the time $t$ evolves,
in Fig. \ref {f3} (a) we plot the effective exponent 
$z(t)$ against $1/t$.  For two dimensions, $z(t)$ starting
from a value around $3$ gradually {\it decreases}
and reaches $2.63(2)$ at $t=640$ (i.e., obtained with data of
$C(r,t)$ at the time $t=640$ and $2t=1280$).
Assuming the behavior of $z(t)$ 
will not changed essentially after $t=1280$,
the extrapolated value of $z$ to the infinite time $t$
is estimated to be $2.6(1)$. Interestingly, for three dimensions
the exponent $z(t)$ starting from a value around $2.5$
{\it increases} slowly, but stabilizes at $2.7$ after
$t=80$. A good estimate of $z$ is $z=2.7(1)$.
Within statistical errors, the values of the dynamic exponent $z$ in 
two and three dimensions coincide with each other,
thus indicating that the dynamic exponent $z$ is dimension independent.
This can also be seen from the joining
of two different curves at relatively larger times
in Fig. \ref {f3} (a).

 \begin{figure}[p]\centering
\epsfysize=6.5cm
\epsfclipoff
\fboxsep=0pt
\setlength{\unitlength}{0.6cm}
\begin{picture}(9,9)(0,0)
\put(-2,-0.5){{\epsffile{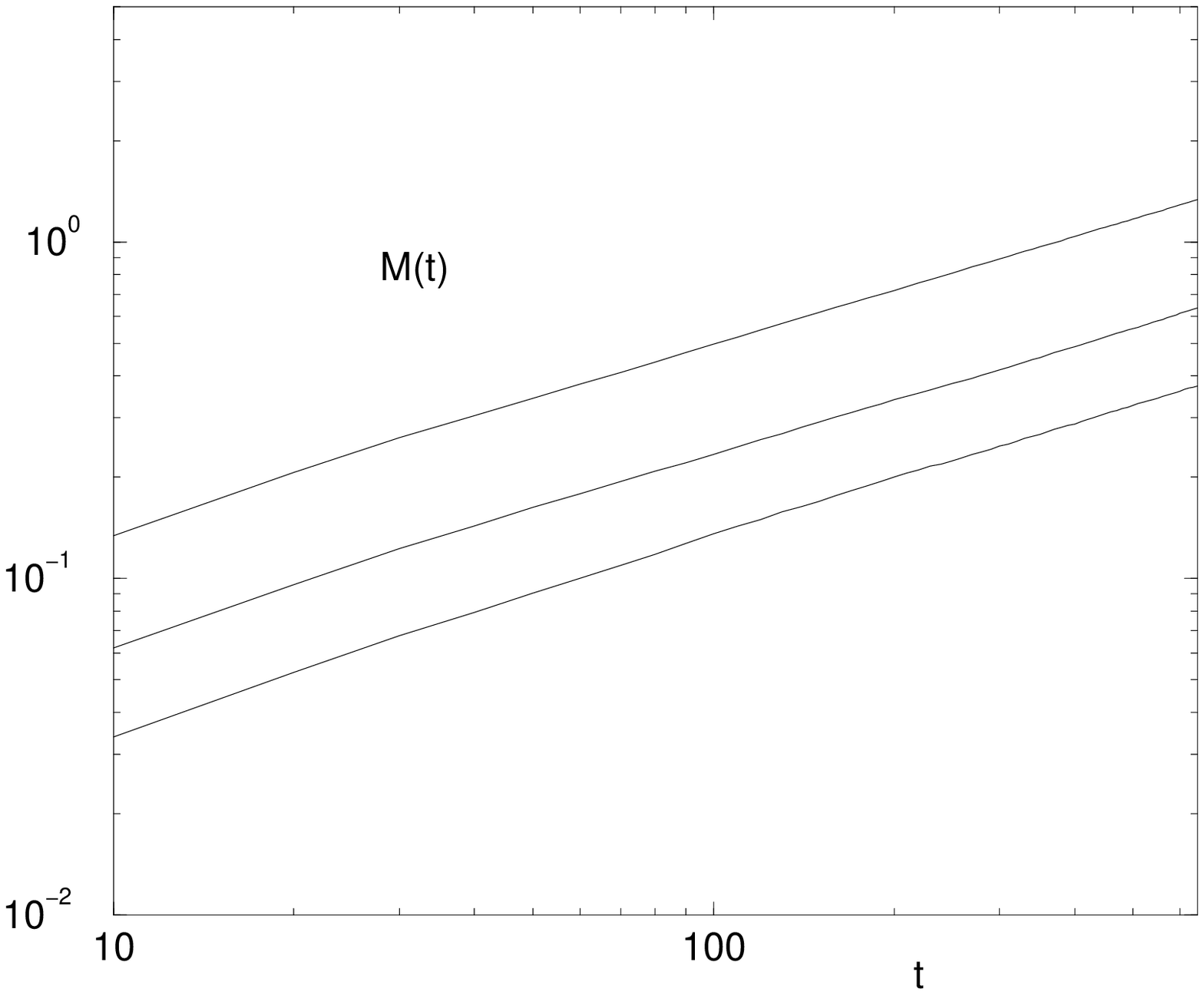}}}
\end{picture}
\caption{The magnetization in three dimensions in log-log scale.
The lattice size is $L=128$. From below,
$m_0=0.00123$, $0.00245$ and $0.00491$.
}
\label{f4}
\end{figure}

In the case of the Ising model with Monte Carlo dynamics,
the effective exponent $z(t)$ in two dimensions converges to
$z=2$ rather fast, e.g., see Ref. \cite {hum91},
but relatively slowly in three dimensions 
due to corrections to scaling. It might be somewhat general
that phase ordering dynamics in three dimensions
is somewhat more complicated than in two dimensions.

An interesting fact observed in Ref. \cite {zhe00b} is that 
even though the dynamic exponent $z$ of the $\phi^4$ theory
 in two dimensions
with Hamiltonian dynamics is different from that of the Ising model
with Monte Carlo dynamics,
the scaling function $f(x)$ in Eq. (\ref {e60}) is the same.
However, this is probably only by chance since it is
{\it not} the case in three dimensions.
The scaling function $f(x)$ of the three dimensional 
$\phi^4$ theory with Hamiltonian dynamics is different not only  from
that of the two dimensional but also from that of the 
three dimensional Ising model with Monte Carlo dynamics. 
The dashed line in Fig. \ref {f2} (a) shows the $f(x)$
of the two-dimensional $\phi^4$ theory.
In general, Hamiltonian dynamics for isolated systems
differs indeed from stochastic dynamics for non-isolated systems.

For simple understanding of the correlation function $C(r,t)$,
one can measure the time-dependent second moment
$M^{(2)}(t)$. The scaling form results
in a power law behavior for $M^{(2)}(t)$ and from the slope
in log-log scale one can estimate the corresponding exponent. 
Such an approach is rather typical and useful in critical dynamics
\cite {zhe98}. It can also be applied in ordering dynamics,
but less efficient. In critical dynamics, in the scaling collapse
of $C(r,t)$ one has to determine two exponents,
 the dynamic exponent $z$ and the static exponent $2 \beta /\nu$.
 Therefore, it is efficient to read out directly 
 the exponent $(d-2 \beta /\nu)/z$
 from the slope of of $M^{(2)}(t)$ in log-log scale
 \cite {zhe98}.
 However, in ordering dynamics the 'static' exponent
 $2 \beta /\nu=0$ and the scaling collapse of $C(r,t)$ is
 only a one parameter fit. Measurements of $M^{(2)}(t)$
 do not show advantage since it is not self-averaged
 and there is larger fluctuation for bigger lattices.
  This is seen from the data
 in Ref. \cite {zhe00b}. Anyway, in Fig. \ref {f2} (b) we have plotted
 the second moment in log-log scale for the three dimensional 
$\phi^4$ theory. The power law behavior is seen after $t \sim 80$
and this is consistent with Fig. \ref {f3} (a).
According to Eq. (\ref {e70}),
the resulting dynamic exponent is $z=2.5(2)$, consistent
within errors with $z=2.7(1)$ measured from $C(r,t)$.

Another interesting exponent in ordering dynamics is 
$\lambda$ governing the power law decay of the auto-correlation
$A(t)$ in Eq. (\ref {e90}). The measurements
of the auto-correlation in ordering dynamics is easier than in
critical dynamics since the fluctuation is much smaller.
The results for the $\phi^4$ theory in two and three dimensions are
shown in Fig. \ref {f1} (b) and \ref {f2} (b).
In order to see how the effective exponent $\lambda/z$ depends on the time 
$t$, we have measured the slope of the curves 
in a time interval $[t, 2t]$.
The results are given in Table \ref {t1}.
For both two and three dimensions, the exponent $\lambda/z$
becomes stable after $t=160$.
The final values are $\lambda/z=0.466(3)$ and $0.618(4)$
for two and three dimensions respectively.
To show clearly that our estimates of $\lambda/z$ are indeed
reasonable,
in Fig. \ref {f3} (b) we plot $A(t) t^a$ as a function of the time $t$.
A correct value $a=\lambda/z$ should result in
a constant for $A(t) t^a$, at least for larger times.
Such a behavior is nicely seen from the lower solid line
and the dashed line for two and three dimensions
in the figure. To confirm that the value $\lambda/z=0.466(3)$
for two dimensions is really different from $\lambda/z=0.625$
for stochastic dynamics, the corresponding curve
with $a=0.625$ is also displayed there (the upper solid line).
 Obviously, it does not tend to a constant.

\begin{table}[h]\centering
\begin{tabular}{cccccc}
  $t$   &  40 &  80      &   160 & 320  &  640 \\
\hline
$2d$ & 0.508(1) & 0.492(1) & 0.469(7)  & 0.461(6) &   0.463(6) \\
\hline
$3d$ & 0.633(4) & 0.609(1) & 0.617(3) &   0.619(7) &            \\
\end{tabular}
\caption{The exponent $\lambda/z$ measured in a time interval
$[t, 2t]$ from the auto-correlation in two and three dimensions.
}
\label{t1}
\end{table}

From measurements of $z$ (from $C(r,t)$) and $\lambda/z$, we estimate
the exponent $\lambda=1.21(5)$ and $1.67(6)$ 
for two and three dimensions respectively.
For stochastic dynamics, theoretical prediction for two
dimensions is $\lambda=1.25$ \cite {fis88,bra94},
but in Monte Carlo simulations it is usually slightly
smaller \cite {hum91}. Extrapolation is needed to obtain 
a value very close to $1.25$. There is always some uncertainty in
extrapolation. Therefore, we tend to claim
that $\lambda$ of the $\phi^4$ theory in two dimensions
with Hamiltonian dynamics 
is the same as that of stochastic dynamics.
In three dimensions, our $\lambda=1.67(6)$ agrees
very well with the 'best' theoretical prediction $1.67$
for stochastic dynamics \cite {liu91,bra94}. Numerical
measurements of $\lambda$ for stochastic dynamics
in three dimensions look somewhat problematic
and the results fluctuate around the theoretical values.

To complete our investigation, we have also 
simulated the initial increase of the magnetization
in Eq. (\ref {e100}). Since the exponent $\theta$ is relatively big, 
compared with that in critical dynamics 
\cite {zhe00b,zhe98}, we need to prepare a very small 
initial magnetization $m_0$.
In Fig. \ref {f4}, the magnetization in three dimensions 
is plotted in log-log scale for
$m_0=0.00123$, $0.00245$ and $0.00491$ (from below)
respectively. The power law behavior is observed after 
$t \sim 50$. From the slope,
we measure the exponent $\theta$.
Within statistical errors, we can not find any $m_0$
dependence of $\theta$. The value of $\theta$ is estimated
to be $0.55(2)$. With $\theta$ in hand, combining
$\lambda/z=0.618(4)$, we obtain another value for the
dynamic exponent, $z=2.6(1)$.

In Table \ref {t2}, all the exponents measured for the 
$\phi^4$ theory with Hamiltonian dynamics are summarized.
Results for two dimensions are taken from Ref. \cite {zhe00b},
but $\lambda/z$, $\lambda$ and $z$ from $C(r,t)$
are slightly modified according to new data in this paper.
Different measurements in two and three dimensions
suggest that $z=2.6(1)$ is a good estimate
for the dynamic exponent. Different from the case of
critical dynamics \cite {zhe98}, the critical exponent
$\theta$ in phase ordering dynamics has not yet drawn enough
attention, even though it has been addressed \cite {bra94}.
One reason might be that in ordering dynamics,
increasing of the magnetization is expected if a non-zero 
initial value $m_0$ is set, but in critical dynamics,
this is anomalous. Anyway, we think $\theta$ is interesting
since it gives another independent estimate for
the dynamic exponent $z$ or $\lambda$.

\begin{table}[h]\centering
\begin{tabular}{c|cc|ccc|c}
         &          &                      &                  &  $z$ 
                    &  &\\
\hline
         & $\theta$ & $\lambda/z$          & $d/(\lambda/z+\theta)$ & $C(r,t)$
                 & $M^{(2)}$     & $\lambda$       \\
\hline
  2d & 0.31(1)    &   0.466(3)   &    2.6(1)    &  2.6(1)
                 &   2.6(1)    &  1.21(5)     \\
\hline
  3d &  0.55(2)   & 0.618(4)     &   2.6(1)       &    2.7(1)
                 &  2.5(2)    &  1.67(6)  \\
\end{tabular}
\caption{Exponents of the $\phi^4$ theory with Hamiltonian dynamics.
To calculate $\lambda$,
$z$ measured from $C(r,t)$ is taken as input.
}
\label{t2}
\end{table}

\section{Conclusions}

In conclusion, we have numerically solved 
the Hamiltonian equations of motion for the two- and
three-dimensional $\phi^4$ theory with random initial states.
Phase ordering dynamics is carefully investigated.
Scaling behavior is confirmed. The dynamic exponent $z$ 
is dimension independent. Different measurements
yield a value $z=2.6(1)$ and it is different from $z=2$ 
for stochastic dynamics of model A. The scaling function
for the equal-time spatial correlation function
is dimension dependent, and in general also
different from that of stochastic dynamics of model A
(it is the same probably only by chance in two dimensions).
However, the exponent $\lambda$ of Hamiltonian dynamics
is the same as that of stochastic dynamics of model A.

\section*{Acknowledgments}
 Work is supported in part by
DFG, TR 300/3-1, and by DFG, GRK 271.

\end{document}